\documentclass[%
 aip,
 amsmath,amssymb,
 reprint,%
 onecolumn,
]{revtex4-1}

\usepackage{graphicx}
\usepackage{dcolumn}
\usepackage{bm}

\usepackage[utf8]{inputenc}
\usepackage[T1]{fontenc}
\usepackage{mathptmx}
\usepackage{etoolbox}
\usepackage{color}

\allowdisplaybreaks

\makeatletter
\def\@email#1#2{%
 \endgroup
 \patchcmd{\titleblock@produce}
  {\frontmatter@RRAPformat}
  {\frontmatter@RRAPformat{\produce@RRAP{*#1\href{mailto:#2}{#2}}}\frontmatter@RRAPformat}
  {}{}
}%
\makeatother
\begin{document}

\title{Various integral estimations and screening schemes for extended systems
in PySCF}

\author{Qiming Sun}
\email{osirpt.sun@gmail.com}
\affiliation{Axiomquant Investment Management LLC, Beijing 100871, China}

\date{\today}

\begin{abstract}
\noindent
In this document, we briefly review the two-electron integral algorithms based
on the range-separated algorithms and Fourier transform integral algorithms that
are implemented in the PySCF package. For each integral algorithm, we estimate
the upper bound of relevant integrals and derive the necessary conditions for
the screening parameters, including distance cutoff and energy cutoff, to reach
the desired accuracy.
Our numerical tests show that the proposed integral estimators and
screening parameters can effectively address the required accuracy
while computational efforts are not wasted on unintended accuracy.

\end{abstract}

\maketitle

\section{Introduction}
Gaussian basis with periodic treatments have been made available for
extended systems to compute Hartree-Fock, density functional theory (DFT) and
post-mean-field methods in various program
packages\cite{Sun2018,Sun2020,WCMS-CP2K,Kuehne2020,Erba2022}.
To characterize periodicity, the Gaussian basis employed by crystalline
calculations requires an infinity of primitive Gaussian functions recurrently placed
in the repeated image cells.
Unlike the integral evaluation program for molecules, evaluating a single
crystalline integral involves the computation of massive
primitive Gaussian integrals.
Thanks to the exponential decay of Gaussian function,
given a specific requirement on numerical precision, many primitive integrals
can be neglected without breaking the translational symmetry of crystalline
integrals.
Proper integral screening schemes for extended systems need to be
developed to filter the negligible primitive integrals.

For Coulomb-type interactions, integral screening relies on an accurate
estimation of electron repulsion integrals (ERI).
Integral estimation has big impact on the accuracy and the computational cost in
crystalline calculations.
Underestimation may cause a loss of accuracy, while overestimation may lead to a
waste of computational efforts.
Schwarz inequality is the simplest while quite useful integral estimator although
it always overestimates the integral value.
By including the factor of distance between charge densities,
improved Schwarz inequality estimators were proposed in the past\cite{Gill1994,Lambrecht2005,Maurer2012,Maurer2013,Hollman2015,Valeev2020}.
They can be used to screen four-center ERIs and density fitting methods for the
bare Coulomb operator and the complementary error function attenuated Coulomb
operator\cite{Izmaylov2006,Thompson2019}.

Integral screening for crystalline integrals is more complex than for
molecular integrals due to the presence of periodicity.
Crystalline integrals can be evaluated with different integral
algorithms\cite{Lippert1999,Carsky2012,Ben2013,Burow2009,Izmaylov2006,Kudin2000,Maschio2008,Pisani2008,Usvyat2007,Varga2006,James2008,Guidon2009,Sun2017,Ye2021,Ye2021a,Ye2022,Sun2020a,Bintrim2022,Sharma2022}.
Various parameters, such as the range of truncated Coulomb
operator, the multipole expansion order, the energy cutoff, the real space
grids, etc. have to be used to efficiently compute integrals.
It often requires preliminary numerical experiments or certain experience to
tune these parameters to achieve desired accuracy without sacrificing performance.
Among the crystalline integral evaluation algorithms,
Ye developed the range-separated Gaussian density fitting (RSDF)
algorithm\cite{Ye2021} and explored the integral estimators and integral
screening scheme for the short-range part of
the integrals required by RSDF. His integral estimator helps RSDF algorithm
gain a magnitude speed-up comparing to the earlier GDF
implementation\cite{Ye2021,Sun2017}.

In this work, we document integral estimators and integral screening parameters for
the crystalline integral algorithms developed in the PySCF package, including
the range-separated density fitting\cite{Ye2021}, the compensated-charge density fitting
(CCDF)\cite{Sun2017}, the range-separated exact exchange algorithm
(RSJK)\cite{Sun2020a} and the Fourier transform integral algorithm.
In Section~\ref{sec:algorithms}, we briefly review these crystalline integral
algorithms.
In Section~\ref{sec:cutoffs}, we discuss the integral estimation and various
cutoffs and screening parameters for different types of integrals.
The effectiveness of integral screening schemes are assessed in Section~\ref{sec:tests}.

\section{Integral algorithms in PySCF}
\label{sec:algorithms}
The periodicity adapted crystalline Gaussian basis is composed of
primitive Gaussian functions recurrently placed in $N$ image cells
characterized by translational shifts $\mathbf{T}$
\begin{equation}
  \phi_\mu^\mathbf{k}(\mathbf{r})
  = \frac{1}{\sqrt{N}} \sum_\mathbf{T} e^{i\mathbf{k}\cdot\mathbf{T}}
  \chi_\mu(\mathbf{r}-\mathbf{T}),
\end{equation}
where $\chi_\mu(\mathbf{r})$ is a primitive Gaussian function centered at
$\mathbf{R}_\mu$ with normalization factor $N_\mu$
\begin{equation*}
  \chi_\mu(\mathbf{r})
  = N_\mu (x-R_{\mu x})^{m_x} (y-R_{\mu y})^{m_y} (z-R_{\mu z})^{m_z} e^{-\alpha_\mu (\mathbf{r} - \mathbf{R}_\mu)^2}.
\end{equation*}
In an {\it ab initio} calculation of a crystalline system, essentially one needs
to build the overlap integrals and integrals of kinetic operator, nuclear
attraction operator, two-electron Coulomb repulsion operator in terms of the
crystalline basis.

The translational symmetry allowed overlap integrals between two crystalline
basis functions are
\begin{gather}
  S_{\mu\nu}^{\mathbf{k}}
  = \sum_{\mathbf{T}}e^{i\mathbf{k}\cdot\mathbf{T}} S_{\mu\nu^\mathbf{T}},
  \label{eq:overlap}
  \\
  S_{\mu\nu^\mathbf{T}}
  = \int \chi_\mu(\mathbf{r}) \chi_\nu(\mathbf{r} - \mathbf{T}) d^3\mathbf{r}.
\end{gather}
The lattice-sum over vector $\mathbf{T}$ can be truncated according to the overlap
between the two primitive functions $S_{\mu\nu^\mathbf{T}}$. We can compute the
kinetic integrals in a similar manner.

Nuclear attraction integrals can be computed using the algorithm for
three-center two-electron integrals, which we will discuss later.
In this treatment, we use a very sharp s-type function to mimic the nuclear
charge distribution
\begin{equation}
  \chi_A(\mathbf{r}) = Z_A \lim_{\zeta\rightarrow\infty} \Big(\frac{\zeta}{2\pi}\Big)^{3/2}
  e^{-\zeta |\mathbf{r}-\mathbf{R}_A|^2}
\end{equation}
and rewrite the integral of nuclear attraction to
\begin{align}
  V_{N,\mu\nu}^{\mathbf{k}}
  = \sum_{A\in \text{cell 0}}
  \sum_{\mathbf{MN}}e^{i\mathbf{k}\cdot(\mathbf{N-M})}
  \int \frac{\chi_\mu(\mathbf{r}_1-\mathbf{M}) \chi_\nu(\mathbf{r}_1 - \mathbf{N}) \chi_A(\mathbf{r}_2)}{r_{12}}
  d^3\mathbf{r}_1 d^3\mathbf{r}_2.
\end{align}

It is relatively straightforward to compute
two-electron Coulomb repulsion integrals with the assistance of plane-waves
\begin{equation}
  \frac{e^{i\mathbf{G}\cdot \mathbf{r}}}{\sqrt{2\pi}}.
\end{equation}
In reciprocal space, the four-center two-electron integrals can be evaluated
\begin{equation}
  g_{\mu\nu,\kappa\lambda}^{\mathbf{k}_\mu\mathbf{k}_\nu\mathbf{k}_\kappa\mathbf{k}_\lambda}
  = \frac{1}{\Omega} \sum_{\mathbf{G}}
  \frac{4\pi\rho_{\mu\nu}^{\mathbf{k}_\mu\mathbf{k}_\nu}(\mathbf{G}+\mathbf{k}_{\mu\nu})
  \rho_{\kappa\lambda}^{\mathbf{k}_\kappa\mathbf{k}_\lambda}(-\mathbf{G}+\mathbf{k}_{\kappa\lambda})
  }{|\mathbf{G}+\mathbf{k}_{\mu\nu}|^2},
  \label{eq:aft:eri}
\end{equation}
\begin{equation}
  \mathbf{k}_{\mu\nu} = -\mathbf{k}_\mu + \mathbf{k}_\nu.
\end{equation}
$\Omega$ is the volume of the unit cell. The plane-wave vector $\mathbf{G}$ is
chosen to be integer multipliers of reciprocal lattice vectors.
The Fourier transformed density (or the product of basis functions)
$\rho(\mathbf{G})$ can be obtained with either analytical Fourier transformation
\begin{equation}
  \rho_{\mu\nu}^{\mathbf{k}_\mu\mathbf{k}_\nu}(\mathbf{G})
  = \sum_{\mathbf{T}} e^{i\mathbf{k}_\nu\cdot\mathbf{T}}
  \int e^{-i(\mathbf{G}-\mathbf{k}_{\mu\nu})\cdot\mathbf{r}}
  \chi_\mu(\mathbf{r}) \chi_\nu(\mathbf{r} - \mathbf{T}) d^3\mathbf{r}
  \label{eq:aft:rhoij}
\end{equation}
or discrete Fourier transformation
\begin{equation}
  \rho_{\mu\nu}^{\mathbf{k}_\mu\mathbf{k}_\nu}(\mathbf{G})
  = \frac{1}{\Omega}\sum_\mathbf{r} e^{-i(\mathbf{G}-\mathbf{k}_{\mu\nu})\cdot\mathbf{r}}
  \phi_\mu^{\mathbf{k}_\mu}(\mathbf{r}) \phi_\nu^{\mathbf{k}_\nu}(\mathbf{r}).
\end{equation}
This algorithm can be viewed as a density fitting method using plane-waves as
auxiliary basis to expand the electron density.
Implementations of this algorithm are available in PySCF with the name FFTDF (fast Fourier
transform density fitting) and AFTDF (analytical Fourier transform density fitting).

Computing the two-electron integrals with Fourier transform is expensive in many
scenario. Thanks to the locality of Gaussian function in either real space or
reciprocal space, recipes that mix real-space and reciprocal-space
integral evaluation were developed. They are the range-separated integral
algorithms RSDF and RSJK, and the charge-compensated integral algorithm CCDF.

\subsection{Range-separated integral algorithms}
Using the error function and its complementary function to split the Coulomb operator
\begin{equation}
  \frac{1}{r_{12}}
  = \frac{\mathrm{erfc}(\omega r_{12})}{r_{12}}
  + \frac{\mathrm{erf}(\omega r_{12})}{r_{12}},
  \label{eq:coul:split}
\end{equation}
we get a short-range (SR) component associated with the complementary error function
and a long-range (LR) operator associated with the error function.
We refer the Coulomb operator with complementary error function to SR because it
decays exponentially in real space.
In contrast, the LR component has a compact distribution in reciprocal space
\begin{equation}
  \frac{4\pi}{G^2} e^{-\frac{G^2}{4\omega^2}}.
  \label{eq:coul:lr}
\end{equation}
We also split the electron density into
a compact part $\rho_c(\mathbf{r})$ and a diffused part $\rho_d(\mathbf{r})$
based on their compactness in real space. Typically, the diffused part of the density
is constructed using smooth Gaussian functions and is expected to be
compact in reciprocal space.
When computing the two-electron repulsion integrals with the RSJK or RSDF algorithm,
\begin{equation}
  \int \frac{\rho(\mathbf{r}_1) \rho(\mathbf{r}_2)}{r_{12}} d^3 \mathbf{r}_1 d^3 \mathbf{r}_2,
\end{equation}
locality is utilized and the integrals are computed in two steps. In the first
step, we compute the SR Coulomb with the compact density analytically in real
space
\begin{equation}
  \int \frac{\rho_{c}(\mathbf{r}_1)\mathrm{erfc}(\omega r_{12})
  \rho_{c}(\mathbf{r}_2)}{r_{12}} d^3 \mathbf{r}_1 d^3 \mathbf{r}_2.
  \label{eq:int2e:sr}
\end{equation}
For RSJK, computing the four-index analytical integrals requires three nested
lattice-sum
\begin{gather}
  g_{\mu\nu,\kappa\lambda}^{\mathbf{k}_\mu\mathbf{k}_\nu\mathbf{k}_\kappa\mathbf{k}_\lambda}
  = \sum_{\mathbf{MNT}}
  e^{i\mathbf{k}_\nu\cdot\mathbf{N}-i\mathbf{k}_\mu\cdot\mathbf{M}-i\mathbf{k}_\kappa\cdot\mathbf{T}}
  g_{\mu^\mathbf{M}\nu^\mathbf{N},\kappa^\mathbf{T}\lambda}
  \label{eq:j4c:triple:lattice:sum}
  \\
  g_{\mu^\mathbf{M}\nu^\mathbf{N},\kappa^\mathbf{T}\lambda}
  = \int \frac{\chi_\mu(\mathbf{r}_1-\mathbf{M}) \chi_\nu(\mathbf{r}_1 - \mathbf{N})
  \mathrm{erfc}(\omega r_{12})\chi_\kappa(\mathbf{r}_2-\mathbf{T}) \chi_\lambda(\mathbf{r}_2)}{r_{12}}
  d^3\mathbf{r}_1 d^3\mathbf{r}_2.
  \label{eq:eri4c2e:rs}
\end{gather}
For RSDF, a single lattice-sum is required for the two-center SR integrals and
a double lattice-sum is required for the three-center SR integrals
\begin{gather}
  g_{\mu,\nu}^{\mathbf{k}}
  = \sum_{\mathbf{N}}
  e^{i\mathbf{k}_\nu\cdot\mathbf{N}}
  g_{\mu,\nu^\mathbf{N}},
  \\
  g_{\mu,\nu^\mathbf{N}}
  = \int \frac{\chi_\mu(\mathbf{r}_1)\mathrm{erfc}(\omega r_{12}) \chi_\nu(\mathbf{r}_2 - \mathbf{N})}{r_{12}}
  d^3\mathbf{r}_1 d^3\mathbf{r}_2,
  \\
  g_{\mu\nu,\kappa}^{\mathbf{k}_\mu\mathbf{k}_\nu}
  = \sum_{\mathbf{MN}}
  e^{i\mathbf{k}_\nu\cdot\mathbf{N}-i\mathbf{k}_\mu\cdot\mathbf{M}}
  g_{\mu^\mathbf{M}\nu^\mathbf{N},\kappa},
  \label{eq:j3c:double:lattice:sum}
  \\
  g_{\mu^\mathbf{M}\nu^\mathbf{N},\kappa}
  = \int \frac{\chi_\mu(\mathbf{r}_1-\mathbf{M}) \chi_\nu(\mathbf{r}_1 - \mathbf{N})
  \mathrm{erfc}(\omega r_{12})\chi_\kappa(\mathbf{r}_2)}{r_{12}}
  d^3\mathbf{r}_1 d^3\mathbf{r}_2.
  \label{eq:j2c:lattice:sum}
\end{gather}
In the second step, rest terms, including the LR Coulomb operator or the
terms with diffused part of electron density, are all collected and evaluated in
reciprocal space numerically, which can be shortly denoted as
\begin{equation}
  \frac{1}{\Omega}\sum_{\mathbf{G}} \frac{4\pi}{G^2}
  \Big(\rho(\mathbf{G})\rho(-\mathbf{G})
  -(1-e^{-\frac{G^2}{4\omega^2}})\rho_c(\mathbf{G})\rho_c(-\mathbf{G})\Big).
  \label{eq:int2e:lr}
\end{equation}
All terms in this step exbibit a compact distribution in reciprocal space,
which allows for rapid truncation of the summation over the plane-wave functions
in the formula above. More details of the four-center ERI algorithm can be
found in reference \onlinecite{Sun2020a}.

\subsection{Charge-compensated algorithms}
In CCDF, we partition the auxiliary basis function into two components, the
zero-multipole component and the plane-wave component
\begin{gather}
  \varphi_\mu^\mathbf{k}(\mathbf{r})
  = \frac{1}{\sqrt{N}} \sum_\mathbf{T} e^{i\mathbf{k}\cdot\mathbf{T}}
  [\chi_\mu(\mathbf{r}-\mathbf{T}) - \xi_\mu(\mathbf{r}-\mathbf{T})]
  + \frac{1}{(2\pi)^3}\sum_{\mathbf{G}} e^{i(\mathbf{G}+\mathbf{k})\cdot\mathbf{r}}
  \rho_{\xi_\mu}(\mathbf{G}+\mathbf{k}).
  \label{eq:cc:auxbas}
\end{gather}
The zero-multipole function $\varphi$ is a regular Gaussian function
compensated by a smooth Gaussian function $\xi$ that has the same charge (or
multipoles). The effect of $\xi$ is eliminated by the plane-wave component
\begin{gather}
  \xi_\mu(\mathbf{r})
  = \frac{N_\mu}{N_\eta}
  (x-R_{\mu x})^{m_x} (y-R_{\mu y})^{m_y} (z-R_{\mu z})^{m_z} e^{-\alpha_\mu (\mathbf{r} - \mathbf{R}_\mu)^2}
  , \quad \eta < \alpha_\mu,
  \\
  \rho_{\xi_\mu}(\mathbf{G})
  = \int e^{-i\mathbf{G}\cdot \mathbf{r}} \xi_\mu(\mathbf{r}) d^3\mathbf{r}.
\end{gather}
For the two-center and the three-center integrals integrals involving $\varphi$,
we carry out the analytical integral scheme in real space
\begin{gather}
  g_{\mu,\nu}^{\mathbf{k}}
  = \sum_{\mathbf{N}}
  e^{i\mathbf{k}_\nu\cdot\mathbf{N}}
  ( J_{\mu,\nu^\mathbf{N}} - J_{\xi_\mu,\nu^\mathbf{N}}
  - J_{\mu,\xi_\nu^\mathbf{N}} + J_{\xi_\mu,\xi_\nu^\mathbf{N}}),
  \\
  J_{\mu,\nu^\mathbf{N}}
  = \int \frac{\chi_\mu(\mathbf{r}_1) \chi_\nu(\mathbf{r}_2 - \mathbf{N})}{r_{12}}
  d^3\mathbf{r}_1 d^3\mathbf{r}_2,
  \\
  g_{\mu\nu,\kappa}^{\mathbf{k}_\mu\mathbf{k}_\nu}
  = \sum_{\mathbf{MN}}
  e^{i\mathbf{k}_\nu\cdot\mathbf{N}-i\mathbf{k}_\mu\cdot\mathbf{M}}
  (J_{\mu^\mathbf{M}\nu^\mathbf{N},\kappa}
  - J_{\mu^\mathbf{M}\nu^\mathbf{N},\xi_\kappa}),
  \label{eq:j3c:ccdf}
  \\
  J_{\mu^\mathbf{M}\nu^\mathbf{N},\kappa}
  = \int \frac{\chi_\mu(\mathbf{r}_1-\mathbf{M}) \chi_\nu(\mathbf{r}_1 - \mathbf{N})
  \chi_\kappa(\mathbf{r}_2)}{r_{12}}
  d^3\mathbf{r}_1 d^3\mathbf{r}_2.
\end{gather}
The integrals associated to the plane-wave component are computed in
reciprocal space. The formula for the two-center two-electron integrals is
\begin{equation}
  \frac{4\pi}{\Omega}\sum_{\mathbf{G}}
  \frac{\rho_\mu^{\mathbf{k}}(\mathbf{G}+\mathbf{k})
  \rho_{\xi_\nu}^{\mathbf{k}}(-\mathbf{G}-\mathbf{k})
  + \rho_{\xi_\mu}^{\mathbf{k}}(\mathbf{G}+\mathbf{k})
  \rho_\nu^{\mathbf{k}}(-\mathbf{G}-\mathbf{k})
  - \rho_{\xi_\mu}^{\mathbf{k}}(\mathbf{G}+\mathbf{k})
  \rho_{\xi_\nu}^{\mathbf{k}}(-\mathbf{G}-\mathbf{k})
  }{|\mathbf{G}+\mathbf{k}|^2}
\end{equation}
and the formula for the three-center two-electron integrals is
\begin{equation}
  \frac{1}{\Omega} \sum_{\mathbf{G}}
  \frac{4\pi\rho_{\mu\nu}^{\mathbf{k}_\mu\mathbf{k}_\nu}(\mathbf{G}+\mathbf{k}_{\mu\nu})
  \rho_\xi(-\mathbf{G}+\mathbf{k}_{\kappa\lambda})
  }{|\mathbf{G}+\mathbf{k}_{\mu\nu}|^2}.
\end{equation}

\section{Cutoffs estimation}
\label{sec:cutoffs}
Based on the position of primitive Gaussian functions, we use distance cutoff
$R_\text{cut}$ to determine which image cells to be included in the lattice-sum.
For plane-wave functions, we use energy cutoff $E_\text{cut}$ to truncation the
summation of plane-waves.

We first show the conversion between $R_\text{cut}$ and the range of lattice-sum.
Let $\Delta\mathbf{R}$ be the displacement between two atoms in the unit cell.
For lattice vectors $\mathbf{a} = (\mathbf{a}_1, \mathbf{a}_2, \mathbf{a}_3$),
$|\mathbf{a}\mathbf{T} + \Delta\mathbf{R}|$ gives the distance between one atom in
the reference cell (cell 0) and another atom in the image cell indicated by
the vector $\mathbf{T}$ (elements of $\mathbf{T}$ are all integers).
Lattice-sum should include all $\mathbf{T}$s which satisfy
\begin{equation}
  |\mathbf{a}\mathbf{T} + \Delta\mathbf{R}| < R_\text{cut}
  \label{eq:rcut:condition}
\end{equation}
Calling QR decomposition for the 3 $\times$ 4 matrix
\begin{equation*}
  \begin{pmatrix}
    \mathbf{a}_1,
    \mathbf{a}_2,
    \mathbf{a}_3,
    \Delta\mathbf{R}
  \end{pmatrix} = \mathbf{q}\mathbf{c},
\end{equation*}
the inequality \eqref{eq:rcut:condition} can be transformed to
\begin{equation*}
  \begin{pmatrix}
    T_x, T_y, T_z, 1
  \end{pmatrix}
  \cdot \mathbf{c}^T \mathbf{c} \cdot
  \begin{pmatrix}
    T_x \\
    T_y \\
    T_z \\
    1
  \end{pmatrix}
  < R_\text{cut}^{2}.
\end{equation*}
This inequality indicates the lower bound and upper bound of $T_z$
\begin{gather}
  T_z^\text{upper} = \mathrm{ceil}\big(\frac{R_\text{cut} - c_{34}}{c_{33}}\big)
  \\
  T_z^\text{lower} = \mathrm{floor}\big(\frac{-R_\text{cut} - c_{34}}{c_{33}}\big)
  \label{eq:rcut2latticesum}
\end{gather}
The lattice-sum along $z$-direction needs to include all integers in the closed
set $[T_z^\text{lower}, T_z^\text{upper}]$.
Similar procedures can be carried out to determine the bound for $T_x$ and $T_y$
as well as the lattice-sum range along $x$-direction and $y$-direction.

Given energy cutoff $E_\text{cut}$ we can determine the minimal number of plane
waves $\mathbf{N}$ in each direction with the inequality
\begin{gather}
  |\mathbf{N} \mathbf{b}|^2 > 2 E_\text{cut}.
\end{gather}
We can find the requirement for $N_z$ with QR decomposition for $\mathbf{b}$
\begin{gather}
  \begin{pmatrix}
    \mathbf{b}_1,
    \mathbf{b}_2,
    \mathbf{b}_3,
  \end{pmatrix} = \mathbf{q}\mathbf{c},
  \\
  N_z \geq \mathrm{ceil}\big(\frac{\sqrt{2E_\text{cut}}}{c_{33}}\big).
  \label{eq:kecut2nz}
\end{gather}
$N_y$ and $N_z$ can be determined in a similar manner.

\subsection{Distance cutoff for Gaussian basis function on real-space grid}
For any grid inside the reference cell, the value of a remote primitive
Gaussian function centered at $\mathbf{R}$ has the upper bound
\begin{equation}
  \chi_\mu(\mathbf{r}-\mathbf{R}) \leq N_\mu |\mathbf{R}|^{l_\mu} e^{-\alpha_\mu |\mathbf{R}|^2}
\end{equation}
We can estimate an overall error $\varepsilon$ due to a single lattice-sum if
neglecting all primitive functions which are placed remoter than $R_\text{cut}$
\begin{equation}
  \varepsilon
  = \sum_{|\mathbf{T}| > R_\text{cut}} \chi_\mu(\mathbf{r}-\mathbf{T})
  \approx \int_{R > R_\text{cut}} \chi_\mu(\mathbf{R}) d^3\mathbf{R}
  < 4 \pi N_\mu \int_{R > R_\text{cut}} R^{l_\mu+2} e^{-\alpha_\mu R^2} dR.
\end{equation}
To ensure the truncation error smaller than the required precision $\tau$,
the value of $R_\text{cut}$ can be determined by solving the inequality
\begin{equation}
  \frac{2 \pi N_\mu R_\text{cut}^{l_\mu+1}
  e^{-\alpha_\mu R_\text{cut}^2}}{\Omega\alpha_\mu}
  < \tau
\end{equation}
To simplify the estimation for basis functions in the same shell which have the
same angular momentum, we approximate the angular part of the normalization
factor
\begin{gather}
  N_\mu \approx N_R(\alpha_\mu,l_\mu) \sqrt{\frac{2l_\mu+1}{4\pi}},
\end{gather}
where $N_R$ is the radial normalization factor
\begin{equation}
  N_R(\alpha_\mu,l_\mu) =
  \sqrt{\frac{2(2\alpha_\mu)^{l_\mu+3/2}}{\Gamma(l_\mu+\frac{3}{2})}}.
\end{equation}

\subsection{Distance cutoff for overlap}
Assuming that $\chi_\mu$ is centered at the coordinate $(0,0,0)$, the overlap
between two primitive functions $\chi_\mu$ and $\chi_\nu$ has an upper limit
\begin{align}
  \langle \chi_\mu|\chi_\nu\rangle
  &\leq N_{\mu} N_{\nu}
  \int x^{l_\mu} (x-|\mathbf{R}_\nu|)^{l_\nu}e^{-\alpha_\mu x^2}
  e^{-\alpha_\nu(x-|\mathbf{R}_\nu|)^2}
  e^{-\alpha_{\mu\nu}y^2} e^{-\alpha_{\mu\nu}z^2} dx dy dz
  \nonumber\\
  &= N_{\mu} N_{\nu} \frac{\pi}{\alpha_{\mu\nu}}
  \int x^{l_\mu} (x-R_\nu)^{l_\nu}e^{-\alpha_\mu x^2} e^{-\alpha_\nu(x-R_\nu)^2} dx.
\end{align}
To simplify the equations, we adopt shorthand notations
\begin{gather}
  \alpha_{\mu\nu} = \alpha_\mu + \alpha_\nu,
  \\
  l_{\mu\nu} = l_\mu + l_\nu,
  \\
  \theta_{\mu\nu} = (\alpha_\mu^{-1} + \alpha_\nu^{-1})^{-1}.
\end{gather}
We then employ the Gaussian product theorem (GPT)
\begin{equation}
  L_{\mu\nu}^{k}(R)
  = \sum_{l=0}^{k}
  \begin{pmatrix}
    l_\mu \\
    l
  \end{pmatrix}
  \begin{pmatrix}
    l_\nu \\
    k-l
  \end{pmatrix}
  \big(\frac{\alpha_\nu R}{\alpha_{\mu\nu}}\big)^{l_\mu-l}
  \big(\frac{-\alpha_\mu R}{\alpha_{\mu\nu}}\big)^{l_\nu+l-k},
  \label{eq:gpt:coeff}
\end{equation}
and derive the primitive overlap integral
\begin{align}
  I_{\mu\nu}
  &= \pi N_{\mu} N_{\nu} e^{-\theta_{\mu\nu}R_\nu^2}
  \sum_{k}^{l_{\mu\nu}} L_{\mu\nu}^{k}(R_\nu)
  \frac{\Gamma(\frac{k+1}{2})}{\alpha_{\mu\nu}^{(k+3)/2}}.
\end{align}
$\Gamma(s) = \Gamma(s, 0)$ is the incomplete gamma function
\begin{equation}
  \Gamma(s, t) = \int_t^\infty x^{s-1} e^{-x} dx.
\end{equation}
By approximating $\Gamma(\frac{k+1}{2}) \approx \sqrt{\pi}$
\begin{equation}
  I_{\mu\nu}
  \lesssim
  \pi N_{\mu} N_{\nu} e^{-\theta_{\mu\nu}R_\nu^2}
  \sum_{k}^{l_{\mu\nu}}|L_{\mu\nu}^{k}(R_\nu)|
  \frac{\sqrt{\pi}}{\alpha_{\mu\nu}^{(k+3)/2}},
\end{equation}
we can factorize the GPT term and obtain the overlap integral upper bound
\begin{equation}
  I_{\mu\nu}
  \lesssim
  N_\mu N_\nu e^{-\theta_{\mu\nu}R^2}
  \Big(\frac{\alpha_\nu R}{\alpha_{\mu\nu}} + \frac{1}{\sqrt{\alpha_{\mu\nu}}}\Big)^{l_\mu}
  \Big(\frac{\alpha_\mu R}{\alpha_{\mu\nu}} + \frac{1}{\sqrt{\alpha_{\mu\nu}}}\Big)^{l_\nu}
  \Big(\frac{\pi}{\alpha_{\mu\nu}}\Big)^{3/2}.
  \label{eq:factorize:ovlp}
\end{equation}
After considering the effect of the single lattice-sum in the overlap integral \eqref{eq:overlap},
we derive an approximate value of the truncation error $\varepsilon$
\begin{align}
  \varepsilon
  &< \sum_{|\mathbf{R}|>R_\text{cut}} I_{\mu\nu}
  \approx \frac{4\pi}{\Omega} \int_{R > R_\text{cut}} R^2 I_{\mu\nu} dR
  \approx \frac{2\pi R_\text{cut}}{\Omega \theta_{\mu\nu}} I_{\mu\nu}
  < \tau.
\end{align}
Solving this inequality for a specific precision requirement $\tau$, we can
obtain $R_\text{cut}$ for the overlap integrals of crystalline basis.

\subsection{Distance cutoff for Fourier transform}
The analytical Fourier transform of Gaussian function product $\chi_\mu
\chi_\nu$ is
\begin{equation}
  \int e^{-i \mathbf{G}\cdot \mathbf{r}} \chi_\mu(\mathbf{r}) \chi_\nu(\mathbf{r}) d^3\mathbf{r}
  = \pi N_{\mu} N_{\nu} e^{-\frac{G^2}{4\alpha_{\mu\nu}}} e^{-\theta_{\mu\nu}R^2}
  \sum_{k}^{l_{\mu\nu}} L_{\mu\nu}^{k}(R)
  \sum_m
  \begin{pmatrix}
    k \\
    m
  \end{pmatrix}
  \frac{(\frac{-iG_x}{2\alpha_{\mu\nu}})^{k-m}
  \Gamma(\frac{m+1}{2})}{\alpha_{\mu\nu}^{(m+3)/2}}.
  \label{eq:aft:rhoij1}
\end{equation}
Compared to the overlap integrals, the Fourier transform introduces a factor
\begin{equation}
  e^{-\frac{G^2}{4\alpha_{\mu\nu}}} (\frac{G}{2\alpha_{\mu\nu}})^{n}
\end{equation}
which has a maximum value at
\begin{equation}
  \max(e^{-\frac{G^2}{4\alpha_{\mu\nu}}} (\frac{G}{2\alpha_{\mu\nu}})^{n})
  = \Big(\frac{n}{2e\alpha_{\mu\nu}}\Big)^{\frac{n}{2}}.
\end{equation}
This value can hardly be larger than 1 on a regular Gaussian basis in routine
calculations. Therefore, it is sufficient to employ the overlap $R_\text{cut}$
estimator for analytical Fourier transform.

\subsection{Distance cutoff in RSDF}
\label{sec:rcut:rsdf}
Density fitting methods require three-center integrals and two-center integrals.
We first consider the SR-ERI for three primitive functions based on the
multipole expansion estimator developed in Ye's work \onlinecite{Ye2021}
\begin{align}
  g_{\mu\nu\kappa}
  &= \int \frac{\chi_\mu(\mathbf{r}_1)\chi_\nu(\mathbf{r}_1)\mathrm{erfc}(\omega r_{12})\chi_\kappa(\mathbf{r}_2)}{r_{12}}
  d\mathbf{r}^3_1 d^3\mathbf{r}_2
  \nonumber\\
  &\lesssim N_\mu N_\nu N_\kappa
  e^{-\theta_{\mu\nu}d_{\mu\nu}^2}
  \sum_{l}^{l_{\mu\nu}}|L_{\mu\nu}^{l}(d_{\mu\nu})|
  \frac{\pi^3 \nu_{l+l_\kappa}(\theta_{\mu\nu\kappa\omega}, R)}{\alpha_{\mu\nu}^{l+3/2}\alpha_\kappa^{l_\kappa+3/2}},
  \label{eq:srj3c:me}
\end{align}
where
\begin{gather}
  \theta_{\mu\nu}
  = (\alpha_\mu^{-1} + \alpha_\nu^{-1})^{-1},
  \\
  \theta_{\mu\nu\kappa\omega}
  = (\alpha_{\mu\nu}^{-1} + \alpha_\kappa^{-1} + \omega^{-2})^{-1},
  \label{eq:def:theta}
  \\
  d_{\mu\nu}
  = |\mathbf{R}_\mu - \mathbf{R}_\nu|,
  \\
  R = |\mathbf{P}_{\mu\nu} - \mathbf{R}_\kappa|,
  \\
  \mathbf{P}_{\mu\nu}
  = \frac{\alpha_\mu \mathbf{R}_\mu + \alpha_\nu \mathbf{R}_\nu}{\alpha_{\mu\nu}}.
  \label{eq:weighted:center}
\end{gather}
The notation $d_{\mu\nu}$ is the bra separation (between the centers of $\chi_\mu$ and
$\chi_\nu$) and $R$ is the bra-ket separation.
The effective potential $\nu_l(\theta,R)$ has an upper bound
\begin{gather}
  \nu_l(\theta,R)
  = \frac{\Gamma(l+\frac{1}{2}, \theta R^2)}{\sqrt{\pi} R^{l+1}}
  \lesssim \frac{(\theta R)^l e^{-\theta R^2}}{\sqrt{\pi\theta} R^2}
  f_l(\theta R^2),
  \\
  f_l(x) = \sum_{k=0}^{l-1} \frac{(2l-1)!!}{(2l-2k-1)!! (2x)^k}.
\end{gather}
Typically, $1 \leq f_{l}\lesssim 2$ when the bra-ket separation $R$ is reasonably large.
Therefore, we can assume $f_l$ a constant. By applying the factorization
similar to the overlap integral \eqref{eq:factorize:ovlp} we obtain the upper bound of the
integral \eqref{eq:srj3c:me}
\begin{align}
  g_{\mu\nu\kappa}
  &\lesssim N_\mu N_\nu N_\kappa
  e^{-\theta_{\mu\nu}d_{\mu\nu}^2}
  \Big(\frac{\pi^2}{\alpha_{\mu\nu}\alpha_{\kappa}}\Big)^{3/2}
  \sum_{k}^{l_{\mu\nu}}
  |L_{l_\mu,l_\nu}^{k}(d_{\mu\nu})|
  \frac{f_{k+l_\kappa} (\theta_{\mu\nu\kappa\omega} R)^{k+l_\kappa} e^{-\theta_{\mu\nu\kappa\omega} R^2}}
  {\sqrt{\pi\theta_{\mu\nu\kappa\omega}} R^2 \alpha_{\mu\nu}^{k}\alpha_{\kappa}^{l_\kappa}}
  \nonumber\\
  &\leq
  \frac{Q_{\mu\nu}(R) N_\kappa f_l e^{-\theta_{\mu\nu\kappa\omega}R^2}}
  {\sqrt{\pi\theta_{\mu\nu\kappa\omega}} R^2}
  \Big(\frac{\pi}{\alpha_{\kappa}}\Big)^{3/2}
  \Big(\frac{\theta_{\mu\nu\kappa\omega}R}{\alpha_{\kappa}}\Big)^{l_{\kappa}}
  \nonumber\\
  &\leq
  \frac{Q_{\mu\nu}(R) N_\kappa f_l e^{-\theta_{\mu\nu\kappa\omega}R^2}}
  {\sqrt{\pi\theta_{\mu\nu\kappa\omega}} R^2}
  \Big(\frac{\pi}{\alpha_{\kappa}}\Big)^{3/2}
  \Big(\frac{\omega^2 R}{\alpha_{\kappa}+\omega^2}\Big)^{l_\kappa},
\end{align}
where
\begin{equation}
  Q_{\mu\nu}(R)
  =N_\mu N_\nu e^{-\theta_{\mu\nu} d_{\mu\nu}^2} \Big(\frac{\pi}{\alpha_{\mu\nu}}\Big)^{3/2}
  \Big(\frac{\alpha_\nu d_{\mu\nu}}{\alpha_{\mu\nu}} + {\frac{\theta_{\mu\nu\kappa\omega}R}{\alpha_{\mu\nu}}}\Big)^{l_\mu}
  \Big(\frac{\alpha_\mu d_{\mu\nu}}{\alpha_{\mu\nu}} + {\frac{\theta_{\mu\nu\kappa\omega}R}{\alpha_{\mu\nu}}}\Big)^{l_\nu}.
  \label{eq:Qij}
\end{equation}
It is worth noting that $\theta_{\mu\nu\kappa\omega}$ is bounded above
\begin{equation}
  \theta_{\mu\nu\kappa\omega} < (\alpha_{\mu\nu}^{-1} + \omega^{-2})^{-1},
\end{equation}
and 
\begin{equation}
  R < 2R_\text{cut}
\end{equation}
because of the distance cutoff which ensures that all primitive Gaussian
functions and their products must be inside the sphere of diameter
$2R_\text{cut}$.
By considering these bounds, we obtain the upper bound of $Q_{\mu\nu}(R)$
\begin{equation}
  Q_{\mu\nu}^\text{u} =
  N_\mu N_\nu e^{-\theta_{\mu\nu} d_{\mu\nu}^2} \Big(\frac{\pi}{\alpha_{\mu\nu}}\Big)^{3/2}
  \Big(\frac{\alpha_\nu d_{\mu\nu}}{\alpha_{\mu\nu}} + \frac{2\omega^2 R_\text{cut}}{\alpha_{\mu\nu}+\omega^2}\Big)^{l_\mu}
  \Big(\frac{\alpha_\mu d_{\mu\nu}}{\alpha_{\mu\nu}} + \frac{2\omega^2 R_\text{cut}}{\alpha_{\mu\nu}+\omega^2}\Big)^{l_\nu}
\end{equation}
as well as the upper bound of $g_{\mu\nu\kappa}$
\begin{equation}
  g_{\mu\nu\kappa} \lesssim
  \frac{e^{-\theta_{\mu\nu\kappa\omega}R^2}}{R^2}
  \frac{Q_{\mu\nu}^\text{u} N_\kappa f_l}{\sqrt{\pi}\omega}
  \Big(\frac{\pi}{\alpha_{\kappa}}\Big)^{3/2}
  \Big(\frac{2\omega^2 R_\text{cut}}{\alpha_{\kappa}+\omega^2}\Big)^{l_\kappa}.
  \label{eq:eri3c:upperbound}
\end{equation}
For SR ERIs, the inequality above offers a more accurate upper bound
estimation than Schwarz inequality.
In the crystalline integral program, we combine the two estimators.
Schwarz inequality is tested first for each
$g_{\mu\nu\kappa}$ because it is simple and fast to compute.
To reduce the cost of the inequality test \eqref{eq:eri3c:upperbound}, we
precompute $Q_{\mu\nu}^\text{u}$ and the intermediate point center
$\mathbf{P}_{\mu\nu}$ then adjust the integral screening threshold for each
basis product $\chi_\mu \chi_\nu$.
Additionally, $\theta_{\mu\nu\kappa}$ can be precomputed and cached as well
because many basis functions have the same exponent and the number of unique
$\theta_{\mu\nu\kappa}$ is limited.
When computing $g_{\mu\nu\kappa}$, we only require the coordinates of
$\mathbf{P}_{\mu\nu}$ and auxiliary basis $\chi_\kappa$ to compute $R^2$.
Then we test if \begin{equation}
\frac{e^{-\theta_{\mu\nu\kappa\omega}R^2}}{R^2}
\end{equation}
is large enough for the adjusted integral screening threshold.

Next we consider the effects of the double lattice-sum in the integral
\eqref{eq:j3c:double:lattice:sum}.
Without loss of generality, we can assume the center of $\chi_\kappa$ being at $\mathbf{0}$.
In terms of the exponent part of the primitive three-center integral \eqref{eq:srj3c:me}
\begin{equation}
  g_{\mu\nu\kappa} \sim e^{-s}, \quad
  s = \theta_{\mu\nu}d_{\mu\nu}^2 + \theta_{\mu\nu\kappa\omega} R^2,
  \label{eq:j3c:asymptotic}
\end{equation}
we can find the asymptotic behaviour for the integral \eqref{eq:j3c:double:lattice:sum}
\begin{equation}
  \sum_{\mathbf{M}\mathbf{N}} g_{\mu^\mathbf{M}\nu^\mathbf{N}\kappa}
  \sim \sum_{\mathbf{N}} \sum_{\mathbf{M}-\mathbf{N}}
  e^{-\theta_{\mu\nu}|\mathbf{R}_{\mu^\mathbf{M}} - \mathbf{R}_{\nu^\mathbf{N}}|^2}
  e^{-\theta_{\mu\nu\kappa\omega} |\mathbf{P}_{\mu^\mathbf{M}\nu^\mathbf{N}}|^2}.
\end{equation}
This suggests that the contribution from the lattice-sum over $\mathbf{M}$
would decay rapidly, and we can focus on the leading contribution from
the lattice sum of $\mathbf{N}$
\begin{equation}
  \sum_{\mathbf{M}\mathbf{N}} g_{\mu^\mathbf{M}\nu^\mathbf{N}\kappa}
  \sim \sum_{\mathbf{N}} \max_{\mathbf{R}_\mu}(
  e^{-\theta_{\mu\nu}|\mathbf{R}_\mu - \mathbf{R}_{\nu^\mathbf{N}}|^2}
  e^{-\theta_{\mu\nu\kappa\omega} |\mathbf{P}_{\mu\nu^\mathbf{N}}|^2}).
\end{equation}
The double lattice-sum in \eqref{eq:j3c:double:lattice:sum} is reduced to
a single lattice-sum.
Assuming that the remotest primitive function $\chi_\nu$ centered at
$\mathbf{R}_\text{cut}$,
approximately the maximum value of $g_{\mu\nu\kappa}$ can be found when the
center of primitive function $\chi_{\mu}$ is chosen at
\begin{equation}
  \mathbf{R}_\mu
  = \frac{\alpha_{\mu\nu}\alpha_\nu - \alpha_\nu\theta_{\mu\nu\kappa\omega}}
  {\alpha_{\mu\nu}\alpha_\nu + \alpha_\mu\theta_{\mu\nu\kappa\omega}} \mathbf{R}_\text{cut}
  \label{eq:Rmu:opt}
\end{equation}
which minimizes the value of $s$ in \eqref{eq:j3c:asymptotic}
\begin{equation}
  s^* = \theta_{\nu\kappa\omega} R_\text{cut}^2, \quad
  \theta_{\nu\kappa\omega} = (\alpha_\nu^{-1} + \alpha_\kappa^{-1} + \omega^{-2})^{-1}
\end{equation}
with $d_{\mu\nu}$ and $R$ chosen at
\begin{gather}
  d_{\mu\nu}
  = \alpha_\nu^{-1} \theta_{\nu\kappa\omega} R_\text{cut},
  \\
  R
  = \theta_{\mu\nu\kappa\omega}^{-1} \theta_{\nu\kappa\omega} R_\text{cut}.
\end{gather}
At this configuration, we derive the upper bound of the primitive integral $g_{\mu\nu\kappa}$
\begin{equation}
  g_{\mu\nu\kappa} \lesssim
  \frac{2^{l_\mu} \pi^{5/2} N_\mu N_\nu N_\kappa e^{-s^*}
  \theta_{\mu\nu\kappa\omega}^{3/2}(\theta_{\nu\kappa\omega}R_\text{cut})^{l_{\mu\nu\kappa}-2}}
  {\alpha_{\mu\nu}^{l_{\mu}+3/2}\alpha_\kappa^{l_\kappa+3/2}\alpha_\nu^{l_\nu}}
  f_{l_{\mu\nu\kappa}}(\theta_{\mu\nu\kappa\omega}^{-1}\theta_{\nu\kappa\omega}^2 R_\text{cut}^2),
\end{equation}
which suggests the distance cutoff estimator for the three-center integral
\eqref{eq:j3c:double:lattice:sum}
\begin{align}
  \varepsilon
  &< \frac{1}{\Omega}\int_{R>R_\text{cut}} g_{\mu\nu\kappa}d^3\mathbf{R}
  \lesssim
  \frac{2\pi R_\text{cut}}{\Omega\theta_{\nu\kappa\omega}} g_{\mu\nu\kappa}
  < \tau.
  \label{eq:j3csr:estimator}
\end{align}

By carrying out a similar analysis for the two-center primitive SR-ERI, we can approximate
its upper bound
\begin{equation}
  g_{\mu\nu}
  \lesssim N_\mu N_\nu \frac{\pi^3 \nu_{l_{\mu\nu}}(\theta_{\mu\nu\omega}, R)}{\alpha_{\mu}^{l+3/2}\alpha_\nu^{l_\nu+3/2}}.
\end{equation}
After considering the lattice summation effect, we get the radial cutoff
estimator for the two-center integral \eqref{eq:j2c:lattice:sum}
\begin{equation}
  \varepsilon
  \lesssim \frac{2\pi^4 N_\mu N_\nu e^{-\theta_{\mu\nu\omega} R_\text{cut}^2}
  (\theta_{\mu\nu\omega} R_\text{cut})^{l_{\mu\nu}-1}}
  {\Omega\sqrt{\pi\theta_{\mu\nu\omega}} \alpha_{\mu}^{l_\mu+3/2}\alpha_\nu^{l_\nu+3/2}}
  < \tau.
\end{equation}

\subsection{Distance cutoff for RSJK}
The four-center primitive SR-ERI has an approximate value
\begin{align}
  g_{\mu\nu\kappa\lambda}
  &= \int \frac{\chi_\mu(\mathbf{r}_1)\chi_\nu(\mathbf{r}_1)
  \mathrm{erfc}(\omega r_{12})\chi_\kappa(\mathbf{r}_2)\chi_\lambda(\mathbf{r}_2)}{r_{12}}
  d\mathbf{r}^3_1 d^3\mathbf{r}_2
  \nonumber\\
  &\lesssim N_\mu N_\nu N_\kappa N_\lambda
  e^{-\theta_{\mu\nu}d_{\mu\nu}^2} e^{-\theta_{\kappa\lambda}d_{\kappa\lambda}^2}
  \sum_{k}^{l_{\mu\nu}} \sum_{l}^{l_{\kappa\lambda}}
  |L_{l_\mu,l_\nu}^{k}(d_{\mu\nu}) L_{l_\kappa,l_\lambda}^{l}(d_{\kappa\lambda})|
  \frac{\pi^3 \nu_{k+l}(\theta_{\mu\nu\kappa\lambda\omega}, R)}
  {\alpha_{\mu\nu}^{k+3/2}\alpha_{\kappa\lambda}^{l+3/2}},
\end{align}
where
\begin{gather}
  \theta_{\mu\nu\kappa\lambda\omega}
  = (\alpha_{\mu\nu}^{-1} + \alpha_{\kappa\lambda}^{-1} + \omega^{-2})^{-1},
  \\
  R = |\mathbf{P}_{\mu\nu} - \mathbf{P}_{\kappa\lambda}|.
\end{gather}
We then derive the approximation of the upper bound for 4-center SR ERIs
\begin{align}
  g_{\mu\nu\kappa\lambda}
  &\lesssim
  \frac{Q_{\mu\nu}(R) Q_{\kappa\lambda}(R) f_l e^{-\theta_{\mu\nu\kappa\lambda\omega}R^2}}
  {\sqrt{\pi\theta_{\mu\nu\kappa\lambda\omega}} R^2}.
\end{align}
This estimator can be combined with Schwarz inequality to screen integrals
in a manner similar to the 3-center integral screening scheme we discussed in
Section~\ref{sec:rcut:rsdf}.

For the triple lattice-sum in the integral \eqref{eq:j4c:triple:lattice:sum}, it
can also be reduced to a single lattice-sum for the similar
reason we analyzed in Section~\ref{sec:rcut:rsdf}.
The lattice-sum of $\mathbf{M}$ and $\mathbf{T}$ in
\eqref{eq:j3c:double:lattice:sum} decays exponentially, and only the lattice-sum
of $\mathbf{N}$ needs to be analyzed.
Asymptotically,
\begin{equation}
  g_{\mu\nu\kappa\lambda} \sim e^{-s}, \quad
  s = \theta_{\mu\nu}d_{\mu\nu}^2+\theta_{\kappa\lambda}d_{\kappa\lambda}^2+\theta_{\mu\nu\kappa\lambda\omega} R^2.
  \label{eq:sr4c:asymptotic}
\end{equation}
Assuming that $\mathbf{R}_\lambda = \mathbf{0}$,
the minimal value of $s$ can be found at
\begin{gather}
  s^* = \theta_{\nu\lambda\omega} R_\nu^2, \quad
  \theta_{\nu\lambda\omega} = (\alpha_\nu^{-1} + \alpha_\lambda^{-1} + \omega^{-2})^{-1},
\end{gather}
when the positions of function $\chi_{\mu}$ and $\chi_\kappa$ are chosen at
\begin{gather}
  \mathbf{R}_\mu
  = \frac{\alpha_{\mu\nu}\alpha_\nu\alpha_\kappa\theta_{\mu\nu\kappa\lambda\omega}
  -\alpha_{\kappa\lambda}\alpha_\nu\alpha_\lambda\theta_{\mu\nu\kappa\lambda\omega}
  +\alpha_{\mu\nu}\alpha_\nu\alpha_{\kappa\lambda}\alpha_\lambda}
  {\alpha_{\mu\nu}\alpha_\nu\alpha_\kappa\theta_{\mu\nu\kappa\lambda\omega}
  +\alpha_{\kappa\lambda}\alpha_\mu\alpha_\lambda\theta_{\mu\nu\kappa\lambda\omega}
  +\alpha_{\mu\nu}\alpha_\nu\alpha_{\kappa\lambda}\alpha_\lambda} \mathbf{R}_\nu,
  \label{eq:Rmu:opt1}
\\
  \mathbf{R}_\kappa
  = \frac{\alpha_{\mu\nu}\alpha_\nu\alpha_{\kappa\lambda}\theta_{\mu\nu\kappa\lambda\omega}}
  {\alpha_{\mu\nu}\alpha_\nu\alpha_\kappa\theta_{\mu\nu\kappa\lambda\omega}
  +\alpha_{\kappa\lambda}\alpha_\mu\alpha_\lambda\theta_{\mu\nu\kappa\lambda\omega}
  +\alpha_{\mu\nu}\alpha_\nu\alpha_{\kappa\lambda}\alpha_\lambda} \mathbf{R}_\nu.
  \label{eq:Rkappa:opt}
\end{gather}
This configuration corresponds to the maximum value of
$g_{\mu\nu\kappa\lambda}$ approximately
\begin{equation}
  g_{\mu\nu\kappa\lambda}
  \lesssim
  \frac{2^{l_{\mu\kappa}} \pi^{5/2} N_\mu N_\nu N_\kappa N_\lambda e^{-s^*}
  \theta_{\mu\nu\kappa\lambda\omega}^{3/2}
  (\theta_{\nu\lambda\omega}R_\text{cut})^{l_{\mu\nu}+l_{\kappa\lambda}-2}}
  {\alpha_{\mu\nu}^{l_{\mu}+3/2}\alpha_{\kappa\lambda}^{l_{\kappa}+3/2}
  \alpha_\nu^{l_\nu}\alpha_\lambda^{l_\lambda}}
  f_{l_{\mu\nu\kappa\lambda}}(\theta_{\mu\nu\kappa\lambda\omega}^{-1} \theta_{\nu\lambda\omega}^2 R_\text{cut}^2).
\end{equation}
We then obtain the requirement of $R_\text{cut}$ for 4-center SR ERIs of RSJK
algorithm
\begin{align}
  \frac{2\pi R_\text{cut}}{\Omega\theta_{\nu\lambda\omega}} 
  \frac{2^{l_{\mu\kappa}} \pi^{5/2} f_{l_{\mu\nu\kappa\lambda}}
  N_\mu N_\nu N_\kappa N_\lambda e^{-s^*}
  \theta_{\mu\nu\kappa\lambda\omega}^{3/2}
  (\theta_{\nu\lambda\omega}R_\text{cut})^{l_{\mu\nu}+l_{\kappa\lambda}-2}}
  {\alpha_{\mu\nu}^{l_{\mu}+3/2}\alpha_{\kappa\lambda}^{l_{\kappa}+3/2}
  \alpha_\nu^{l_\nu}\alpha_\lambda^{l_\lambda}}
  < \tau.
  \label{eq:srj4c:me}
\end{align}

\subsection{Distance cutoff for CCDF}
A regular three-center ERI is approximately\cite{Ye2021,Hollman2015,Valeev2020}
\begin{equation}
  J_{\mu\nu,\kappa}
  \approx N_\mu N_\nu N_\kappa e^{-\theta_{\mu\nu}d_{\mu\nu}^2}
  \sum_{l}^{l_{\mu\nu}} L_{l_\mu,l_\nu}^{l}(d_{\mu\nu})
  \frac{\pi^3(\Gamma(l+l_\kappa+\frac{1}{2}) - \Gamma(l+l_\kappa+\frac{1}{2}, \theta_{\mu\nu\kappa}R^2))}
  {\alpha_{\mu\nu}^{l+3/2}\alpha_\kappa^{l_\kappa+3/2}\sqrt{\pi} R^{l+l_\kappa+1}} .
\end{equation}
As shown in Eq. \eqref{eq:j3c:ccdf},
the compensated function $\chi_\xi$ and the auxiliary function $\chi_\kappa$ are
combined when evaluating the analytical three-center integrals
\begin{equation}
  J_{\mu\nu,\kappa} - J_{\mu\nu,\xi}
  \sim \frac{\Gamma(l+l_\kappa+\frac{1}{2}, \theta_{\mu\nu\eta}R^2) -
  \Gamma(l+l_\kappa+\frac{1}{2}, \theta_{\mu\nu\kappa}R^2)}{R^{l+l_\kappa+1}},
  \label{eq:ccdf:j3c:asymptotic}
\end{equation}
where
\begin{gather}
  \theta_{\mu\nu\kappa} = (\alpha_{\mu\nu}^{-1} + \alpha_\kappa^{-1})^{-1},
  \\
  \theta_{\mu\nu\eta} = (\alpha_{\mu\nu}^{-1} + \eta^{-1})^{-1}.
\end{gather}
In CCDF algorithm, we always have $\theta_{\mu\nu\eta} < \theta_{\mu\nu\kappa}$
because the function $\chi_\xi$ is chosen to be the most smooth function.
For sufficiently large $R$, the second $\Gamma$ function in Eq.
\eqref{eq:ccdf:j3c:asymptotic} is negligible
\begin{equation}
  J_{\mu\nu,\kappa} - J_{\mu\nu,\xi}
  \lesssim N_\mu N_\nu N_\kappa e^{-\theta_{\mu\nu}d_{\mu\nu}^2}
  \sum_{l}^{l_{\mu\nu}} L_{l_\mu,l_\nu}^{l}(d_{\mu\nu})
  \frac{\pi^3\Gamma(l+l_\kappa+\frac{1}{2}, \theta_{\mu\nu\eta}R^2)}
  {\alpha_{\mu\nu}^{l+3/2}\eta^{l_\kappa+3/2}\sqrt{\pi} R^{l+l_\kappa+1}}.
\end{equation}
Analysis similar to Section~\ref{sec:rcut:rsdf} can be carried out, which suggests
the $R_\text{cut}$ estimator of the three center integrals \eqref{eq:j3c:ccdf}
for CCDF
\begin{gather}
  \frac{2^{l_\mu+1} \pi^{7/2} N_\mu N_\nu N_\kappa e^{-s^*}
  \theta_{\mu\nu\eta}^{3/2}(\theta_{\nu\eta}R_\text{cut})^{l_{\mu\nu\kappa}-2}R_\text{cut}}
  {\Omega\alpha_{\mu\nu}^{l_{\mu}+3/2}\alpha_\kappa^{l_\kappa+3/2}\alpha_\nu^{l_\nu}\theta_{\nu\eta}}
  f_{l_{\mu\nu\kappa}}(\theta_{\mu\nu\eta}^{-1}\theta_{\nu\eta}^2 R_\text{cut}^2)
  < \tau
  \label{eq:ccdf:j3c:me}
\end{gather}
where
\begin{gather}
  s^* = \theta_{\nu\eta} R_\text{cut}^2,
  \\
  \theta_{\nu\eta} = (\alpha_{\nu}^{-1} + \eta^{-1})^{-1}.
\end{gather}

\subsection{Energy cutoff for four-center Coulomb integrals}
The error of a two-electron ERI due to energy cutoff $E_\text{cut}$
can be estimated
\begin{equation}
  \varepsilon(E_\text{cut}) = 
  \frac{1}{\Omega} \sum_{|\mathbf{G}|^2>2E_\text{cut}} \frac{4\pi}{G^2}
  \rho_{\mu\nu}(\mathbf{G}) \rho_{\kappa\lambda}(-\mathbf{G})
  <
  16 \pi^2 \int_{\sqrt{2E_\text{cut}}}^\infty \rho_{\mu\nu}(G) \rho_{\kappa\lambda}(G) d G.
  \label{eq:ecut:error4c2e}
\end{equation}
Based on Eq. \eqref{eq:aft:rhoij1} the Fourier transform for orbital products, we
obtain the leading term of $\rho_{\mu\nu}(G)$
\begin{equation}
  \rho_{\mu\nu}(G)
  = |\rho_{\mu\nu}(\mathbf{G})|
  \approx N_{\mu} N_{\nu} e^{-\frac{G^2}{4\alpha_{\mu\nu}}} e^{-\theta_{\mu\nu}d_{\mu\nu}^2}
  (\frac{G}{2\alpha_{\mu\nu}})^{l_{\mu\nu}} 
  (\frac{\pi}{\alpha_{\mu\nu}})^{3/2}
\end{equation}
Given energy cutoff $E_\text{cut}$, the largest error for a density distribution
$\rho_{\mu\nu}$ comes with the interaction between $\rho_{\mu\nu}$ and the most
compact density
$\rho_{\kappa\kappa}$
\begin{align}
  \varepsilon(E_\text{cut})
  &< 16\pi^2 \int_{\sqrt{2E_\text{cut}}}^\infty \rho_{\mu\nu}(G)
  \rho_{\kappa\kappa}(G) d G
  \nonumber\\
  &\approx 
  \frac{16\pi^2 N_\mu N_\nu \theta_{\mu\nu\kappa\kappa} e^{-\theta_{\mu\nu}d_{\mu\nu}^2}}
  {(2l_\kappa-1)!!(2\alpha_{\mu\nu})^{l_{\mu\nu}}(4\alpha_\kappa)^{2l_\kappa}}
  \big(\frac{\pi^2}{2\alpha_{\mu\nu}\alpha_\kappa}\big)^{3/2}
  (2E_\text{cut})^{(l_{\mu\nu}+2l_\kappa-1)/2}
  e^{-\frac{E_\text{cut}}{2\theta_{\mu\nu\kappa\kappa}}}.
\end{align}
For the entire system, the energy cutoff error can be derived in terms of the
interactions between $\rho_{\kappa\kappa}$ and itself
\begin{equation}
  \varepsilon(E_\text{cut})
  < 16\pi^2 \int_{\sqrt{2E_\text{cut}}}^\infty \rho_{\kappa\kappa}^2(G) d G.
\end{equation}
This error estimation then leads to an inequality of $E_\text{cut}$ with respect
to the required precision $\tau$
\begin{equation}
  8\pi^2 N_\kappa^4(\frac{\pi}{2\alpha_\kappa})^3
  (\frac{E_\text{cut}}{8\alpha_\kappa^2})^{2l_\kappa-1/2} e^{-\frac{E_\text{cut}}{2\alpha_\kappa}}
  < \tau.
  \label{eq:kecut:4c}
\end{equation}

It should be noted that the $E_\text{cut}$ error estimation above is derived
with the assumption that the Fourier transform of
$\rho_{\mu\nu}(\mathbf{G})$ is analytically computed.
If $\rho_{\mu\nu}(\mathbf{G})$ are computed with the fast Fourier transform (FFT)
algorithm on $N$ discrete real-space grids
\begin{equation}
  \rho_{\mu\nu}(\mathbf{G}) \sim \frac{\Omega}{N}
  \sum_{n}^N e^{-i \mathbf{G} \cdot \mathbf{r}_n}
  \phi_\mu^*(\mathbf{r}_n) \phi_\nu(\mathbf{r}_n),
\end{equation}
the $E_\text{cut}$ estimation \eqref{eq:kecut:4c} is not enough because the error
of FFT was not considered.
When working on FFT two-electron integrals,
one also needs to ensure that the Fourier transform for the orbital product
is converged tightly to an error smaller than the required precision.
The FFT electron density is
\begin{equation}
  \mathrm{FFT}[\rho(\mathbf{r})]
  = \frac{\Omega}{N}\sum_{n=0}^N e^{-i\mathbf{G}\cdot\mathbf{r}_n}\rho(\mathbf{r}_n),
  \quad |\mathbf{G}| \leq \sqrt{2E_\text{cut}}.
\end{equation}
We can transform and split the electron density $\rho(\mathbf{r})$ according to
the momentum of plane-waves
\begin{align}
  \rho(\mathbf{r})
  &=\frac{1}{\Omega}\sum_{|\mathbf{G}|=0}^{\infty}
  e^{i\mathbf{G}\cdot\mathbf{r}_n}\rho(\mathbf{G})
  \\
  &= \frac{1}{N}\sum_{G\leq\sqrt{2E_\text{cut}}}
  e^{i\mathbf{G}\cdot\mathbf{r}}\rho(\mathbf{G})
  + \frac{1}{N}\sum_{G>\sqrt{2E_\text{cut}}}
  e^{i\mathbf{G}\cdot\mathbf{r}}\rho(\mathbf{G}).
\end{align}
The error of FFT electron density is thereby around
\begin{align}
  \mathrm{FFT}[\rho(\mathbf{r})] - \rho(\mathbf{G})
  &=\Big(\frac{1}{N}\sum_{G'}^\infty\sum_{n=0}^N  e^{-i\mathbf{G}\cdot\mathbf{r}_n}
  e^{i\mathbf{G}'\cdot\mathbf{r}_n}\rho(\mathbf{G}')\Big)
  -\rho(\mathbf{G})
  \\
  &=\sum_{G'>\sqrt{2E_\text{cut}}}\frac{1}{N}\sum_{n=0}^N
  e^{-i\mathbf{G}\cdot\mathbf{r}_n}
  e^{i\mathbf{G}'\cdot\mathbf{r}_n}\rho(\mathbf{G}')
  \\
  &\lesssim\sum_{G'>\sqrt{2E_\text{cut}}} \rho(\mathbf{G}').
\end{align}
The error for FFT two-electron integrals can be approximated
\begin{equation}
  \varepsilon
  \approx\frac{1}{\Omega}\sum_{|\mathbf{G}|=0}^{\sqrt{2E_\text{cut}}}
  V(\mathbf{G})[\mathrm{FFT}[\rho(\mathbf{r})] - \rho(\mathbf{G})]
  \lesssim v \sum_{G'>\sqrt{2E_\text{cut}}} \rho(\mathbf{G}'),
\end{equation}
where
\begin{equation}
  v = \frac{1}{\Omega}\sum_{|\mathbf{G}|=0}^{\sqrt{2E_\text{cut}}} V(\mathbf{G}).
\end{equation}
In practice, we find that the energy cutoff for nuclear attraction integrals
\eqref{eq:ecut:nuc} is enough to converge the FFT two-electron integrals.

\subsection{Energy cutoff for three-center Coulomb integrals}
The Fourier transform for a single Gaussian function is
\begin{equation}
  \rho_\kappa(\mathbf{G})
  = \int e^{-i \mathbf{G}\cdot \mathbf{r}} \chi_\kappa(\mathbf{r}) d^3 \mathbf{r}
  = \pi N_\kappa e^{-\frac{G^2}{4\alpha_\kappa}}
  \sum_k
  \begin{pmatrix}
    l_\kappa \\
    k
  \end{pmatrix}
  \frac{(\frac{-iG_x}{2\alpha_\kappa})^{l_\kappa-k}\Gamma(\frac{k+1}{2})}{\alpha_\kappa^{(k+3)/2}}.
\end{equation}
For Coulomb interactions between $\rho_{\mu\nu}(G)$
and $\rho_\kappa(G)$, we can derive the  $E_\text{cut}$ error,
\begin{align}
  \varepsilon(E_\text{cut})
  &< 16\pi^2 \int_{\sqrt{2E_\text{cut}}}^{\infty} \rho_{\mu\nu}(G) \rho_{\kappa}(G) dG
  \nonumber\\
  &\approx\frac{32\pi^2N_\mu N_\nu N_\kappa e^{-\theta_{\mu\nu}d_{\mu\nu}^2}}
  {(2\alpha_{\mu\nu})^{l_{\mu\nu}-1}(2\alpha_\kappa)^{l_\kappa}}
  \Big(\frac{\pi^2}{\alpha_{\mu\nu}\alpha_\kappa}\Big)^{\frac{3}{2}}
  (2E_\text{cut})^{\frac{l_{\mu\nu}+l_\kappa-1}{2}}
  e^{-\frac{E_\text{cut}}{2\theta_{\mu\nu\kappa}}}
  < \tau,
  \label{eq:ecut:error3c2e}
\end{align}
where
\begin{equation*}
  \theta_{\mu\nu\kappa}
  = (\alpha_{\mu\nu}^{-1} + \alpha_\kappa^{-1})^{-1}.
\end{equation*}

\subsection{Energy cutoff for nuclear attraction integrals}
\label{sec:nuc}
When calculating nuclear attraction integrals, we can use steep s-type
Gaussian functions to mimic the charge distribution of point nuclear charges
\begin{equation}
  \chi_\kappa
  = \sum_A \lim_{\zeta\rightarrow\infty}\Big(\frac{\pi}{\zeta}\Big)^{3/2}
  Z_A e^{-\zeta |r-R_A|^2}.
\end{equation}
The three-center energy cutoff analysis \eqref{eq:ecut:error3c2e} is ready to
estimate $E_\text{cut}$ for nuclear attraction integrals with the setting
$l_\kappa=0$ and $\alpha_\kappa\rightarrow\infty$
\begin{equation}
  \varepsilon(E_\text{cut})
  \lesssim\frac{32\pi^2N_\mu N_\nu e^{-\theta_{\mu\nu}d_{\mu\nu}^2}}
  {(2\alpha_{\mu\nu})^{l_{\mu\nu}-1}}
  \Big(\frac{\pi}{\alpha_{\mu\nu}}\Big)^{\frac{3}{2}}
  (2E_\text{cut})^{\frac{l_{\mu\nu}-1}{2}}
  e^{-\frac{E_\text{cut}}{2\alpha_{\mu\nu}}} < \tau.
  \label{eq:ecut:nuc}
\end{equation}

\subsection{Energy cutoff for LR integrals}
For LR integrals, energy cutoff is primarily determined by the Gaussian factor
in the LR Coulomb kernel \eqref{eq:coul:lr}. Similar to the case of
the full Coulomb kernel, we only need to consider the most compact orbital
products in the system to estimate $E_\text{cut}$.
By carrying out the analysis discussed in the previous sections, we obtain the
truncation error as well as the $E_\text{cut}$ inequality for the four-center LR
integrals
\begin{equation}
  \varepsilon(E_\text{cut}) < 
  \frac{32\pi^2 \theta_{\mu\omega} (2E_\text{cut})^{2l_\mu-1}}{((4l_\mu-1)!!)^2}
  e^{-\frac{E_\text{cut}}{2\theta_{\mu\omega}}} < \tau
  \label{eq:ecut:lr4c2e}
\end{equation}
where
\begin{equation}
  \theta_{\mu\omega} = (\alpha_\mu^{-1} + \omega^{-2})^{-1} 
\end{equation}
To ensure $((4l_\mu-1)!!$ has a meaningful value for all angular momentum
$l_{\mu}$, the convention $(-1)!! = 1$ is assumed.

For the three-center LR integrals, the $E_\text{cut}$ estimator can be derived in
a similar fashion
\begin{equation}
  \varepsilon(E_\text{cut}) < 
  \frac{32\pi^2 \theta_{\mu\mu\kappa\omega} 2^{l_\kappa+3/4} (2E_\text{cut})^{l_\mu+(l_\kappa-1)/2}}
  {(4l_\mu-1)!!\sqrt{(4l_\kappa-1)!!}}
  \Big(\frac{\pi}{\alpha_\kappa}\Big)^{\frac{3}{4}}
  e^{-\frac{E_\text{cut}}{2\theta_{\mu\mu\kappa\omega}}} < \tau.
  \label{eq:ecut:lr3c2e}
\end{equation}

\section{Numerical tests and discussion}
\label{sec:tests}
\subsection{Distance cutoff for overlap integrals}
\begin{table}
  \centering
  \caption{Relative error for overlap integral distance cutoff estimation}
  \label{tab:ovlp:estimation}
  \begin{tabular}{llllllll}
    \hline
    $l_\mu$ & $l_\nu$ & $\alpha_\mu = \alpha_\nu$ & $\alpha_\mu = 2\alpha_\nu$ & $\alpha_\mu = 5\alpha_\nu$ & $\alpha_\mu = 100\alpha_\nu$ \\
    \hline
      0 & 0 & 0     & 0     & 0     & 0     \\
      0 & 1 & 0.003 & 0.005 & 0.007 & 0.028 \\
      0 & 2 & 0.006 & 0.009 & 0.015 & 0.058 \\
      0 & 3 & 0.009 & 0.013 & 0.022 & 0.093 \\
      0 & 4 & 0.012 & 0.017 & 0.029 & 0.134 \\
      1 & 0 & 0.003 & 0.002 & 0.001 & 0     \\
      1 & 1 & 0.007 & 0.007 & 0.009 & 0.026 \\
      1 & 2 & 0.010 & 0.011 & 0.015 & 0.053 \\
      1 & 3 & 0.012 & 0.015 & 0.022 & 0.084 \\
      1 & 4 & 0.015 & 0.019 & 0.028 & 0.120 \\
      2 & 0 & 0.006 & 0.005 & 0.003 & 0     \\
      2 & 1 & 0.010 & 0.009 & 0.010 & 0.025 \\
      2 & 2 & 0.012 & 0.013 & 0.016 & 0.050 \\
      3 & 0 & 0.009 & 0.007 & 0.004 & 0.001 \\
      3 & 1 & 0.012 & 0.011 & 0.011 & 0.024 \\
      3 & 2 & 0.015 & 0.015 & 0.017 & 0.048 \\
      4 & 4 & 0.022 & 0.024 & 0.030 & 0.102 \\
    \hline
  \end{tabular}
\end{table}
During the deviations for $R_\text{cut}$ and $E_\text{cut}$,
the factorization approximation \eqref{eq:factorize:ovlp} is widely applied
almost in every integral.
To measure the effectiveness of this
approximation, we compared $R_\text{cut}$ estimated by the overlap estimator
\eqref{eq:factorize:ovlp} to the
precise $R_\text{cut}$ which is solved by a bisection search for the exact
overlap integrals.
Given angular momentum for bra and ket, we noticed that the relative
error for $R_\text{cut}$
only depends on the ratio between the Gaussian exponents of bra and ket.
Table \ref{tab:ovlp:estimation} summarizes the relative errors for various types
of Gaussian basis functions. When the two basis functions have similar shapes (exponents
ratio $<$ 5), the $R_\text{cut}$ errors are small (typically less than 3\%).
However, high angular momentum can slightly increase the error.
When bra and ket have very different shapes, the errors can increase to around 10\%.
Nevertheless, the factorization approximation provides a good estimation for
$R_\text{cut}$ in overlap integrals.

\subsection{Errors for ERIs}
In an SCF calculation, the error of distance cutoff and energy cutoff
estimations may be influenced by several factors, such as the basis set,
size of unit cell, k-point mesh, Coulomb attenuation parameters.
To evaluate the impact of these factors on the cutoff estimations, we computed
ERIs with the range-separated algorithms and compared them
to the benchmark data generated with the reciprocal-space formula
\eqref{eq:aft:eri} with very tight accuracy requirements ($\tau=10^{-16}$).

Unless otherwise specified in each individual test, the test system has one
$s$-type primitive Gaussian function with exponent $\alpha=1.0$ inside a cubic
cell with the edge length $a=1.5$. The Coulomb attenuation parameter for
range-separated algorithms is set to $\omega=0.5$.  Gamma point is adopted for
the integral computation.
In the range-separated algorithm setups, we solve $R_\text{cut}$ and
$E_\text{cut}$ for various precision requirements ($\tau=10^{-5}$ to
$\tau=10^{-12}$) then transform $R_\text{cut}$ and $E_\text{cut}$ to lattice-sum
range and plane-wave summation range using the transformation equations
\eqref{eq:rcut2latticesum} and \eqref{eq:kecut2nz}.
They are used in the triple lattice-sum for the short-range part Eq.
\eqref{eq:j4c:triple:lattice:sum} and the summation over plane-waves for the
long-range part Eq. \eqref{eq:int2e:lr}.

\begin{figure}[htp]
  \centering
  \includegraphics[width=\textwidth]{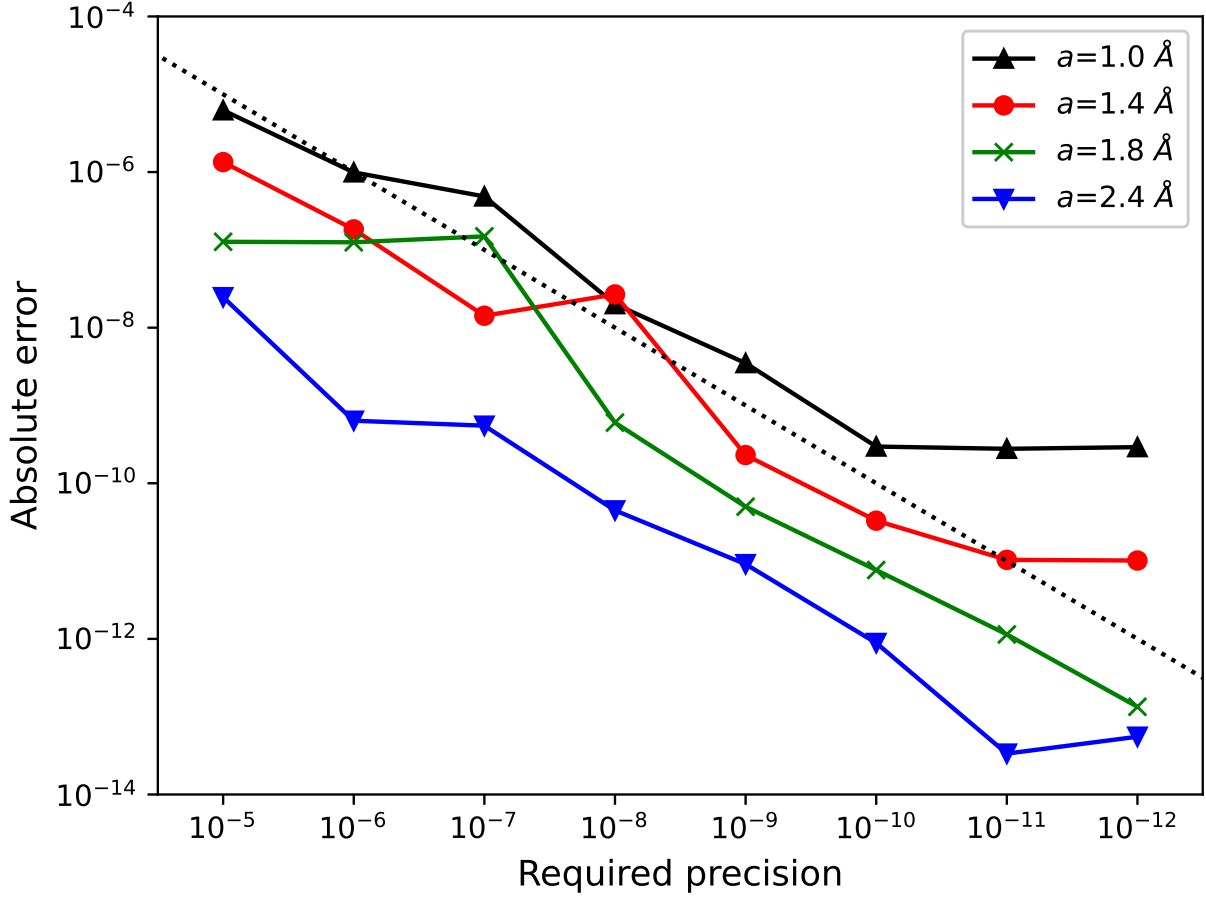}
  \caption{Accuracy tests for lattice parameters}
  \label{fig:eri:a}
\end{figure}

\begin{figure}[htp]
  \centering
  \includegraphics[width=\textwidth]{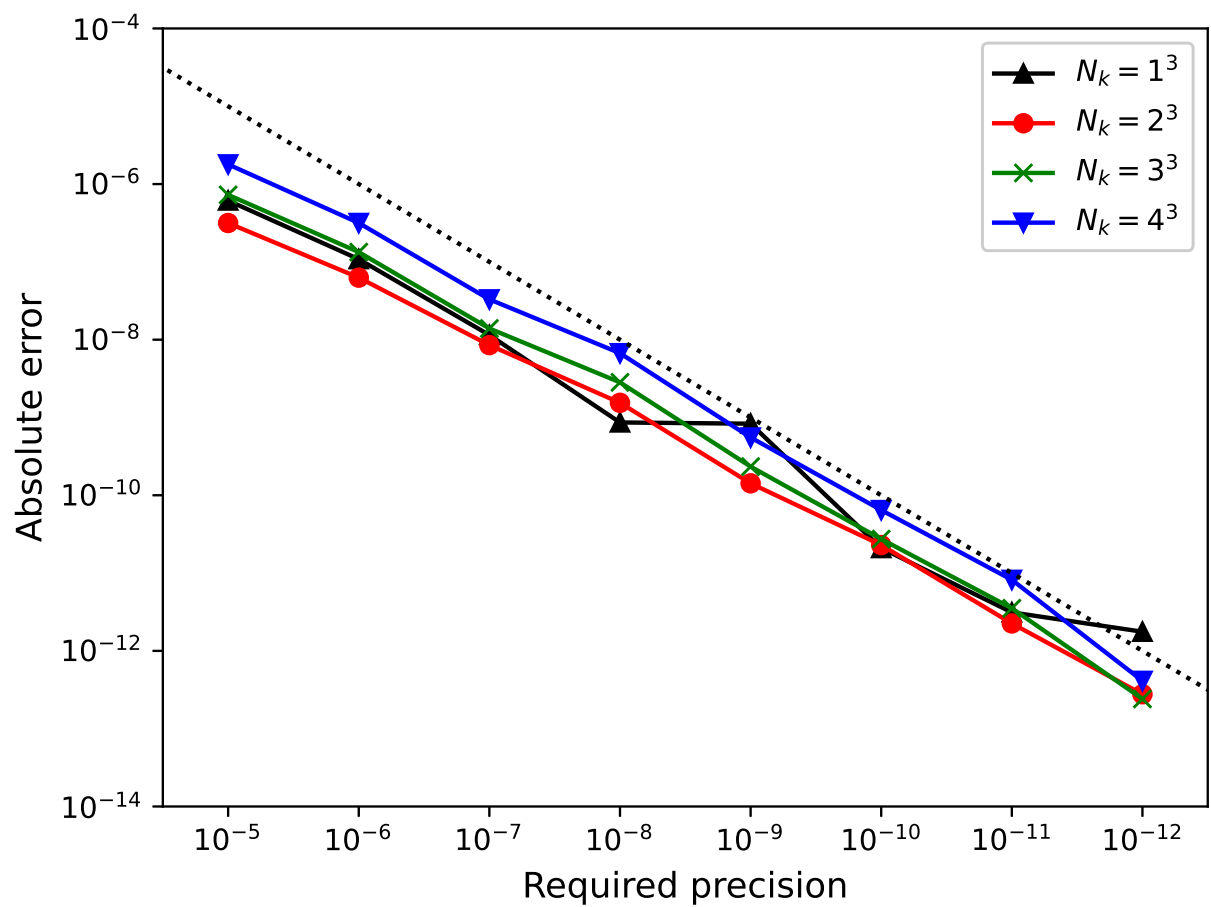}
  \caption{Accuracy tests for k-points}
  \label{fig:eri:kpts}
\end{figure}

\begin{figure}[htp]
  \centering
  \includegraphics[width=\textwidth]{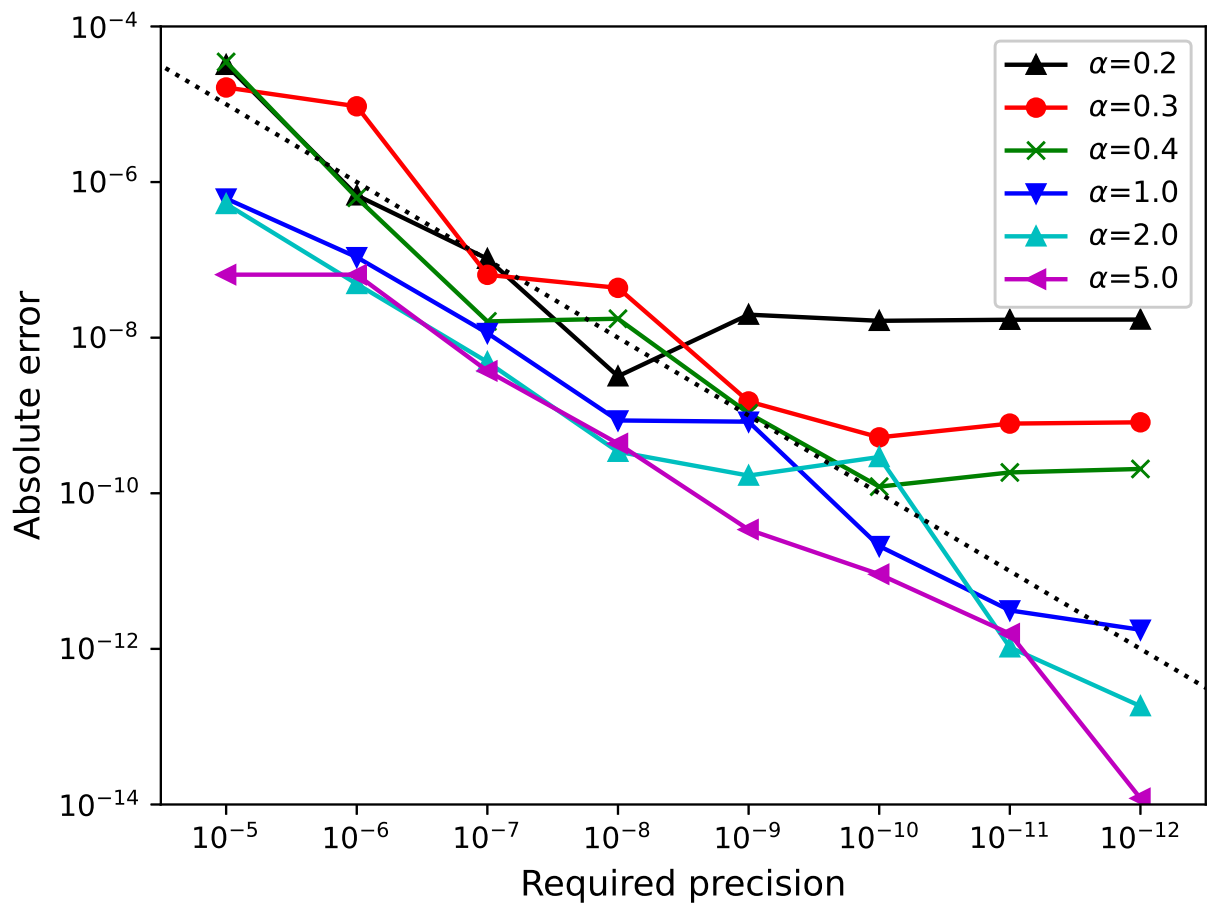}
  \caption{Accuracy tests for exponents of Gaussian basis}
  \label{fig:eri:exp}
\end{figure}

\begin{figure}[htp]
  \centering
  \includegraphics[width=\textwidth]{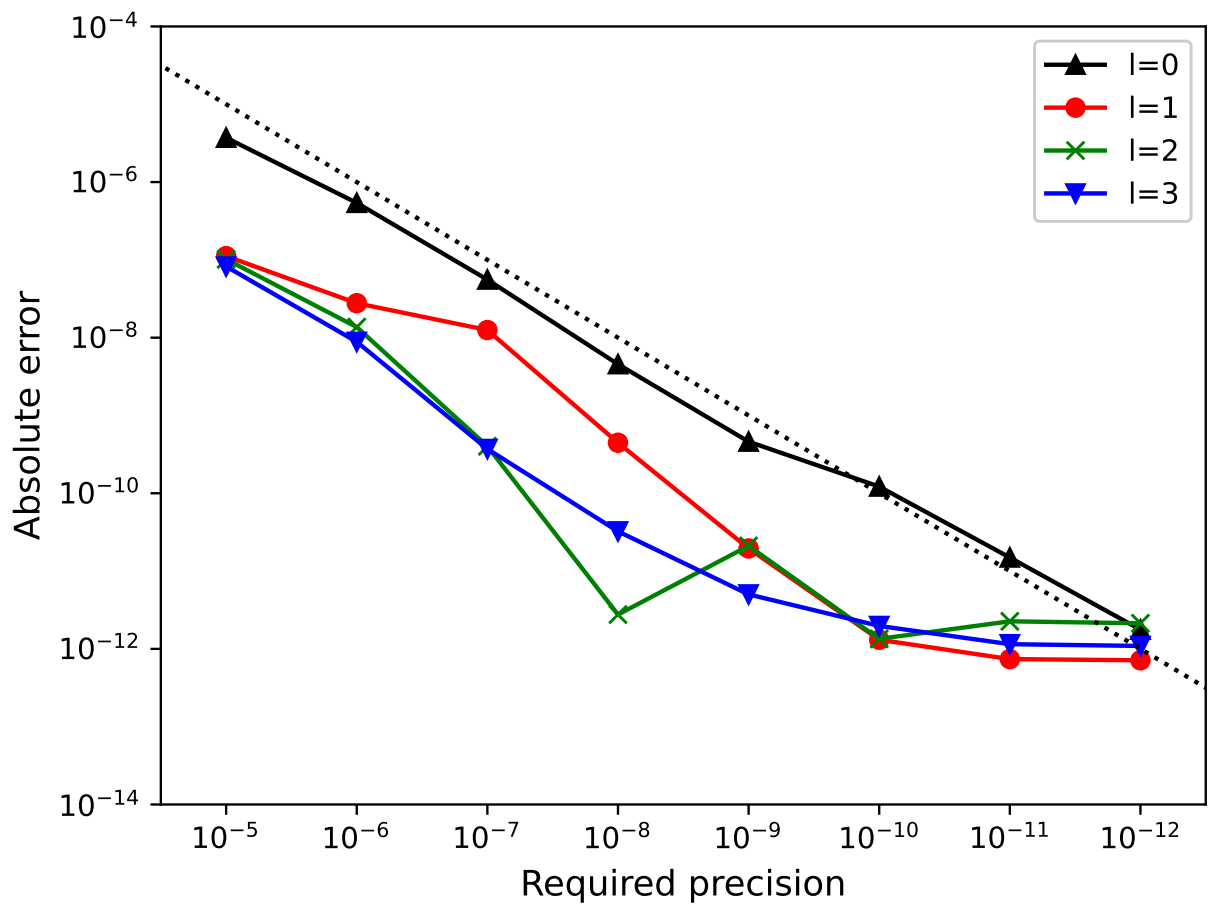}
  \caption{Accuracy tests for angular momentum of Gaussian basis}
  \label{fig:eri:l}
\end{figure}

\begin{figure}[htp]
  \centering
  \includegraphics[width=\textwidth]{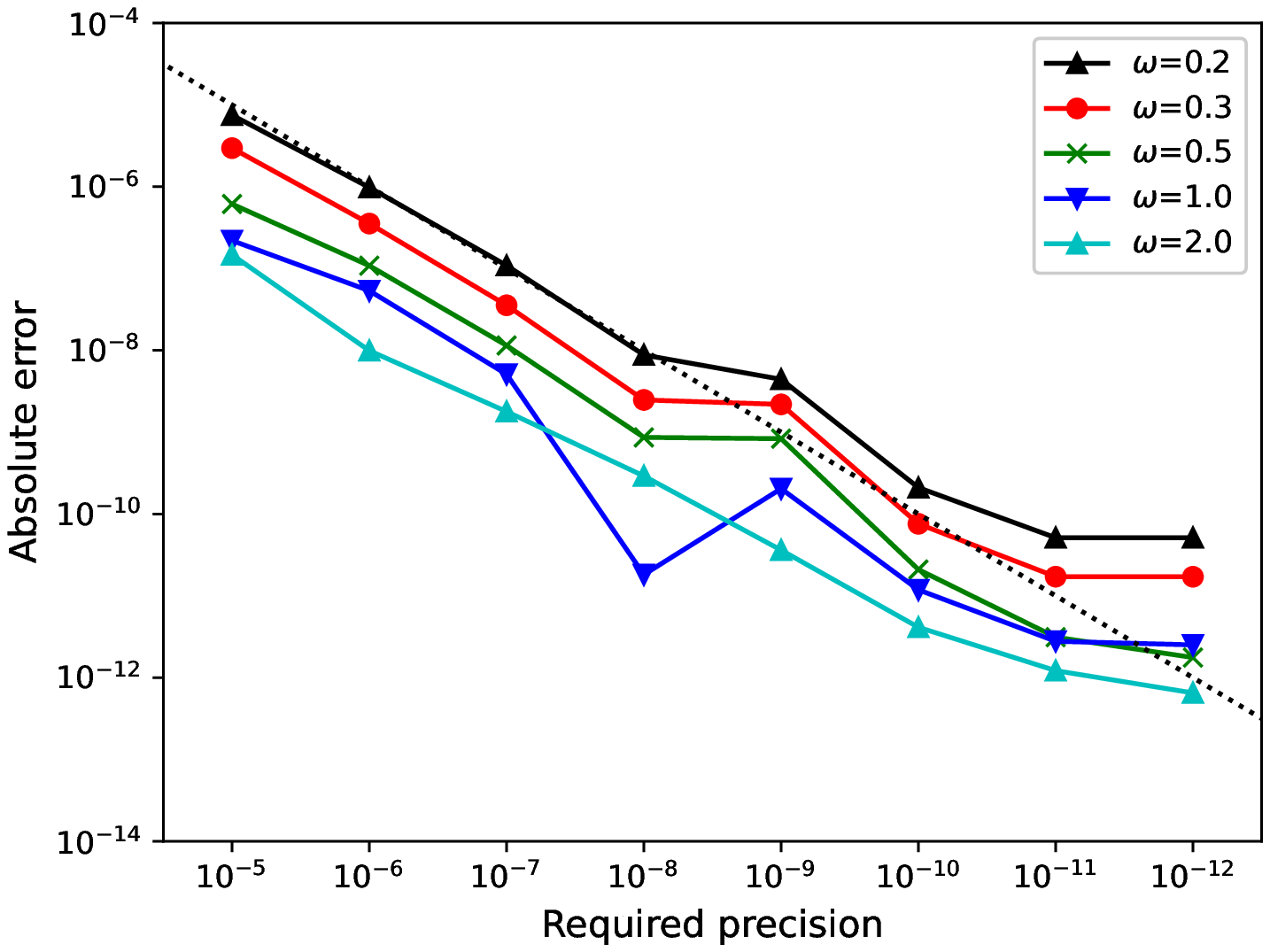}
  \caption{Accuracy tests for Coulomb attenuation parameters}
  \label{fig:eri:omega}
\end{figure}

We found that the accuracy is well-controlled in most tests in the sense that
errors are reduced to a value near or slightly under the desired accuracy as
we increase the precision requirements. It indicates that computational efforts
are being properly utilized without being wasted on the unintended accuracy.
Error underestimation is only observed in a few difficult configurations.
\begin{itemize}
  \item The impact of lattice parameters is exhibited in Figure \ref{fig:eri:a}.
    In this test, we changed the cell edge length from $a=1.0$~\AA~ to $a=2.5$~\AA.
    Generally, small cells lead to larger errors than big cells.
    Good accuracy can be achieved with moderate precision settings up to $10^{-10}$.
    When the required precision is tighter than $10^{-11}$, one may only achieve
    $10^{-10}$ accuracy for the cell with edge length $a=1.0$~\AA.
    One possible reason of the error is the numerical uncertainties in the
    underlying integral library\cite{Sun2015}.
    To confirm that the error is not caused by approximations in cutoff estimators,
    we manually increased the value of $R_\text{cut}$ and $E_\text{cut}$ and
    found that the accuracy was not improved with larger values of $R_\text{cut}$
    or $E_\text{cut}$.
    For the system $a=1.0$ at $\tau=10^{-11}$, the triple lattice-sum in Eq.
    \eqref{eq:j4c:triple:lattice:sum} involves about $1600^3$
    primitive SR-ERIs. Errors in individual primitive integrals, even the
    round-off error, can easily be accumulated to the magnitude around
    $10^{-10}$.

  \item The k-point factor. Integrals
    $(\phi_\mu^{\mathbf{k}_1}\phi_\mu^{\mathbf{k}_2}|\phi_\mu^{\mathbf{k}_2}\phi_\mu^{\mathbf{k}_1})$
    are computed for k-point grids $N_k=1^3 \dots 4^3$ (Figure \ref{fig:eri:kpts}).
    We find similar accuracy performance in all k-point test cases.
    All calculations with various accuracy specifications can reach the required
    accuracy.

  \item Basis effects. We first tested the effects of basis function compactness
    by varying the Gaussian function exponent from $\alpha=0.2$ to $\alpha=5.0$
    (Figure \ref{fig:eri:exp}).
    For relatively compact basis functions, errors are reduced normally as we
    tighten the precision requirements. For diffused basis functions this trend
    only holds up to the precision around $10^{-9}$.
    Errors may also be attributed to the numerical uncertainties in primitive
    integrals. For the basis function with $\alpha=0.2$ at $\tau=10^{-10}$,
    $1500^3$ primitive SR-ERIs have to be included in the triple lattice-sum.

    In the test for basis angular momentum effects, basis functions with angular
    momentum $l=0\dots3$ are tested (Figure \ref{fig:eri:l}). The results show
    that the accuracy is manageable in all test cases, indicating that the
    angular momentum of a basis function is not a significant factor in the
    $R_\text{cut}$ and $E_\text{cut}$ estimation.

  \item In Figure \ref{fig:eri:omega}, we show the errors for Coulomb
    attenuation parameters ($\omega=0.2$ to $\omega=2.0$).
    Similar to the trends we found in lattice parameter tests and basis function
    compactness tests, errors caused by small $\omega$s are larger than errors
    of larger $\omega$s.
    For small $\omega$s, accuracy is limited around $10^{-10}$ because
    of numerical uncertainties.
    Cutoffs are slightly overestimated for large omega.
\end{itemize}



\section{Conclusions}
In this work, we provide a comprehensive analysis of the integral upper
bound and cutoff estimators for the integral algorithms implemented in PySCF.
The distance and energy cutoff estimators derived from the upper bound
estimation are shown to be accurate enough to achieve the required accuracy for
the range-separated integral algorithms while ensuring that computational
resources are efficiently utilized.
Our numerical tests show that the estimators are stable and reliable for various
factors in routine crystalline calculations, such as k-point meshs, basis sets,
unit cell sizes, and Coulomb attenuation parameters.
Uncertainties around $10^{-9}$ may be encountered in certain cases when
involving diffused basis functions, small unit cells, or small Coulomb
attenuation parameters.
Based on the integral estimation derived in this work, we expect that more
aggressive optimization for crystalline integral programs can be carried out.
Comprehensive algorithm and integral screening schemes will need to be designed.
Additional technical details will be considered in future work.

\clearpage
\bibliography{main}

\end{document}